\documentstyle[12pt]{article}
\begin{document}
\begin{center}
{\large \bf
Testing the mechanisms proposed in 
different models for single-spin asymmetries \\ }
\vspace{5mm}
C. Boros, \underline{Z. Liang}, T. Meng and R. Rittel
\\
\vspace{5mm}
{\small\it
Institut f\"{u}r Theoretische Physik der FU Berlin, 
Arnimallee 14, 14195 Berlin, Germany
\\ }
\end{center}

\begin{center}
ABSTRACT\\[-0.3cm]
\vspace{5mm}
\begin{minipage}{130mm}
\small
It is shown that the origin of
the striking left-right asymmetries observed in
single-spin inclusive hadron production processes
in high energy hadron-hadron collisions
can be traced by performing suitable experiments.
Several new experiments are proposed.
The possible outcomes are summarized together with those
of other relevant experiments which have already been
suggested. 
It is pointed out  that the results of such a set of
experiments will not only be able to differentiate between
existing theoretical models, but also be helpful in
locating the source(s) of the observed asymmetries.
\end{minipage}
\end{center}

This is a short summary of 
the main ideas and the main results of a recent paper [1], 
which is available as a preprint since March this year, 
and was briefly summarized in the HERA workshop. 

Recently, much attention has been attracted to 
the study on left-right asymmetries ($A_N$) 
for hadron production
in single-spin high energy hadron-hadron collisions,     
both experimentally [2] and theoretically [3-10]. 
In contrast to the situation at the beginning of 90's, 
we have now already many --- perhaps too many different 
mechanisms which can lead to  
such non-zero $A_N$'s 
in the framework of quantum chromodynamics (QCD) and/or 
quark or quark-parton models.
Hence, we think it is not only useful 
but also necessary to ask:
Can we differentiate these different 
models by performing suitable experiments?

In [1], we showed that this is possible. 
The reasons are the following: 
The available competing theoretical models 
can be mainly divided into the 
following two classes 
(1) perturbative QCD based hard scattering models [3-8] and 
(2) non-perturbative quark-fusion models [10]. 
In the former, 
the cross section for inclusively produced hadrons 
in hadron-hadron collisions 
is expressed as convolution of  
(i) the matrix element for the QCD 
elementary hard scattering process
(ii) the momentum distributions of 
the constituents of the colliding hadrons; and
(iii) the fragmentation function of the scattered 
constituent into the observed hadrons. 
The asymmetry can originate from any one or more of 
these three factors.
In the latter, hadrons in the fragmentation regions 
are products of quark-antiquark fusions and  
the asymmetries originate from the orbital motion 
of the valence quarks in transversely polarized nucleons 
and the hadronic surface effect in single-spin 
hadron-hadron collisions.
The basic ingredients of these models 
are all involved in hadron production 
in hadron-hadron collisions. 
But they may not be involved in other reactions.
This makes it possible to test them individually 
by performing suitable experiments. 
For this purpose, we choose the following 
set of experiments:

\vspace{0.5cm}
\vbox{
\noindent
{\bf Table:} Predictions for left-right asymmetries 
in the discussed experiments 
if the asymmetries observed in inclusive hadron production 
in hadron-hadron collisions originate 
from the different kinds of effects mentioned in the text.

{\footnotesize
\begin{centering}
\begin{tabular}{c|c|c|c|c}
\hline 
&
\multicolumn{4}{c}{
\begin{minipage}[t]{8.cm}
\ \\[-0.3cm]
\phantom {XXXXXX} \hskip -2.cm 
If $A_N$ observed in $p(\uparrow)+p\to \pi +X$ 
 originates from ... 
\end{minipage}}\\[0.2cm]
\begin{minipage}[t]{5.6cm}
\begin{center}
  process \phantom{xxxx}
\end{center}
\end{minipage}
& 
\begin{minipage}[t]{2.cm}
\begin{center}
quark distribution function
\end{center}
\end{minipage}
& 
\begin{minipage}[t]{2.cm}
\begin{center}
elementary scattering process 
\end{center}
\end{minipage} 
&
\begin{minipage}[t]{2.cm}
\begin{center}
quark fragmentation function  
\end{center}
\end{minipage}
& 
\begin{minipage}[t]{2.cm}
\begin{center}
orbital motion \&
surface effect \\ \
\end{center}
\end{minipage} 
\\[0.9cm]  \hline 
\begin{minipage}[t]{5.6cm}
\begin{center}
\ \\[-0.1cm]
 $l+p(\uparrow)\to l+ 
 \left(\begin{array}{cc}\pi^\pm\\ K^+\\ \end{array}\right) + X $
\end{center}
\end{minipage} &  
\begin{minipage}[t]{2.cm}
\begin{center}
\ \\[-0.3cm]
$A_N=0$ \\ wrt jet axis 
\end{center}
\end{minipage}
& 
\begin{minipage}[t]{2.cm}
\begin{center}
\ \\[-0.3cm]
$A_N=0$ \\ wrt jet axis 
\end{center}
\end{minipage}
& 
\begin{minipage}[t]{2.cm}
\begin{center}
\ \\[-0.3cm]
$A_N\not =0$ \\ wrt jet axis 
\end{center}
\end{minipage}
& 
\begin{minipage}[t]{2.cm}
\begin{center}
\ \\[-0.3cm]
$A_N=0$ \\  wrt jet axis
\end{center}
\end{minipage}
\\ [-0.1cm] \cline{2-5}  
\begin{minipage}[t]{5.6cm}
\begin{center}
\ \\[-0.3cm]
 in the current fragmentation region \\ 
 for large $Q^2$ and large $x_B$ 
\end{center}
\end{minipage} &  
\begin{minipage}[t]{2.cm}
\begin{center}
\ \\[-0.3cm]
$A_N\neq 0$ \\ wrt $\gamma^\star$ axis
\end{center}
\end{minipage}
& 
\begin{minipage}[t]{2.cm}
\begin{center}
\ \\[-0.3cm]
$A_N=0$ \\ wrt $\gamma^\star$ axis
\end{center}
\end{minipage}
& 
\begin{minipage}[t]{2.cm}
\begin{center}
\ \\[-0.3cm]
$A_N\not =0$ \\ wrt $\gamma^\star$ axis
\end{center}
\end{minipage}
& 
\begin{minipage}[t]{2.cm}
\begin{center}
\ \\[-0.3cm]
$A_N=0$ \\ wrt $\gamma^\star$ axis
\end{center}
\end{minipage}
\\[0.6cm] \hline 
\begin{minipage}[t]{5.6cm}
\begin{center}
\ \\[-0.1cm]
 $l+ p(\uparrow) \to l+
 \left(\begin{array}{cc} \pi^\pm\\ K^+\\ \end{array}\right) + X $\\ 
 in the target fragmentation region \\     
 for large $Q^2$ and large $x_B$ 
\end{center}
\end{minipage} 
&  
\begin{minipage}[t]{2.cm}
\begin{center}
\ \\[0.4cm]  $A_N\neq 0$
\end{center}
\end{minipage}
& 
\begin{minipage}[t]{2.cm}
\begin{center}
\ \\[0.4cm] $A_N=0$ 
\end{center}
\end{minipage}
& 
\begin{minipage}[t]{2.cm}
\begin{center}
\ \\[0.4cm] $A_N\neq 0$ 
\end{center}
\end{minipage}
& 
\begin{minipage}[t]{2.cm}
\begin{center}
\ \\[0.4cm] $A_N=0$ 
\end{center}
\end{minipage}
\\[1.6cm] \hline 
\begin{minipage}[t]{5.6cm}
\begin{center}
\ \\[-0.1cm]
 $p + p(\uparrow)\to 
\left(\begin{array}{cc} l\bar{l}\\ W^\pm\\ \end{array}\right)+ X $\\
 in the fragmentation region of $p(\uparrow)$
\end{center}
\end{minipage} 
&  
\begin{minipage}[t]{2.cm}
\begin{center}
\ \\[0.2cm] $A_N\neq 0$ 
\end{center}
\end{minipage}
& 
\begin{minipage}[t]{2.cm}
\begin{center}
\ \\[0.2cm] $A_N\approx 0$ 
\end{center}
\end{minipage}
& 
\begin{minipage}[t]{2.cm}
\begin{center}
\ \\[0.2cm] $A_N=0$ 
\end{center}
\end{minipage}
& 
\begin{minipage}[t]{2.cm}
\begin{center}
\ \\[0.2cm] $A_N\neq 0$ 
\end{center}
\end{minipage}
\\ [1.3cm] \hline 
\end{tabular} 
\end{centering} }
}

(A) Perform $l+p(\uparrow )\to l+\pi+X$
for large $x_B$ ($>0.1$, say) and large
$Q^2$ ($>10$ GeV$^2$, say)
and measure the left-right asymmetry
in the current fragmentation region
{\it with respect to the jet axis}.
Here, $l$ stands for charged lepton $e^-$ or  $\mu^-$;
$x_B\equiv Q^2/(2P\cdot q)$ is the usual Bjorken-$x$,
$Q^2\equiv -q^2$, and $P, k, k', q\equiv k-k'$ are the
four momenta of the proton, incoming lepton, outgoing lepton
and the exchanged virtual photon respectively.
The transverse momenta of the produced hadrons
with respect to this axis
come solely from the fragmentation of the quark.
Hence, by measuring this transverse momentum distribution,
we can directly find out whether the fragmentation function
of this polarized quark is asymmetric.

(B) Perform the same kind of experiments as
that mentioned in (A) and
measure the left-right asymmetry of the produced pions
in the current fragmentation region
{\it with respect to the photon direction} in the
rest frame of the proton, and examine those events
where the lepton plane is perpendicular to the
polarization axis of the proton.
In such events, the obtained asymmetry should
contain the contributions
from the intrinsic transverse motion of quarks
in the polarized proton and
those from the fragmentation of polarized quarks,
provided that they indeed exist.

(C) Perform the same kind of experiments as that in (A),
but measure the left-right asymmetry
{\it in the target fragmentation region}
with respect to the moving direction of the proton
in the collider (e.g. HERA) laboratory frame.
By doing so, we are looking at the fragmentation products
of ``the rest of the proton'' complementary to the struck quark
(from the proton).
Since there is no
contribution from the elementary
hard scattering processes and there is no hadronic
surface effect, $A_N$
should be zero if the existence of left-right asymmetries is due to
such effects.
But, if such asymmetries originate from the fragmentation
and/or from the intrinsic transverse motion of the
quarks in the polarized proton,
we should also be able to see them here.

   (D) Measure the left-right asymmetry $A_N$ for $l\bar l$ and/or that for
   $W^\pm$ in $p(\uparrow )+p(0)\to l\bar l \ \mbox {or } W^\pm +X$.
   Here, there is no contribution
   from the quark fragmentation.
   Hence, non-zero values for $A_N$ in such processes can only
   originate from asymmetric quark distributions ---
   including those due to  orbital motion of valence quarks
   and surface effect.
   
  The results of different models for these experiments
  are summarized in the table.
  
 Last but not least, I would like to mention that Professor 
 S.B. Nurushev has informed us, at this symposium, that 
 they have discussed the possibility of carrying out such 
 experiments at HERA in connection with HERA-N project, 
 which will be discussed by W.D. Nowak in another session.

\vspace{0.2cm}
{\small\begin{description}
\itemsep=-0.08truecm
\item{[1]} C. Boros, Z. Liang, T. Meng and R. Rittel, FU-Berlin
           preprint, FUB-HEP/96-4(1996); hep-ph/9603364,   
           also, in Proceedings of the Workshop on the 
           Future Physics at HERA, Sept. 1995 to May 1996. 
\item{[2]} S.~Saroff et al., Phys. Rev. Lett. {\bf 64}, 995 (1990);
          V.D. Apokin et al., Phys. Lett. {\bf B243}, 461 (1990);
  FNAL E581/704 Collab.,
  D.L. Adams et al., Phys. Lett. {\bf B261}, 201 (1991);
  FNAL E704 Collab.,
  D.L.~Adams et al., Phys. Lett. {\bf B264}, 462 (1991); 
   and {\bf B276}, 531 (1992); Z. Phys. {\bf C56},181 (1992);
  A. Yokosawa, in
  {\it Frontiers of High Energy Spin Physics},
   Proceedings of the 10th
   International Symposium, Nagoya, Japan 1992,
   edited by T. Hasegawa {\it et al.}
   (Universal Academy, Tokyo, 1993);
  A.Bravar et al., Phys. Rev. Lett. {\bf 75}, 3073 (1995);  
  {\bf 77}, 2626 (1996); 
  A. Bravar, in these Proceedings. 
\item{[3]} G.~Kane, J.~Pumplin and W.~Repko,
   Phys. Rev. Lett. {\bf 41}, 1689 (1978).
\item{[4]} D. Sivers, Phys. Rev. {\bf D41},83 (1990) and 
             {\bf D41}, 261 (1991).
\item{[5]} J.~Qiu and G.~Sterman, Phys. Rev. Lett. {\bf 67}, 2264 (1991);
          A.~Sch\"afer,  L.~Mankiewicz,  
             P.~Gornicki and S.~G\"ullenstern, 
             Phys. Rev. {\bf D47}, 1 (1993).
\item{[6]} J.~Collins, Nucl. Phys. {\bf B394}, 169 (1993); 
             and {\bf 396}, 161 (1993); 
       J.~Collins, S.F. Hepplelmann and G.A. Ladinsky, 
            Nucl. Phys. {\bf B420}, 565 (1994).
\item{[7]} A.V.~Efremov, V.M.~Korotkiyan, and O.V.~Teryaev,
      Phys. Lett. {\bf B 348}, 577 (1995).
\item{[8]} M.~Anselmino, M.~Boglione, and F.~Murgia, 
       Phys. Lett.  {\bf B362}, 164 (1995).
\item{[9]} S.M. Troshin and N.E. Tyurin, 
             Phys. Rev. {\bf D52}, 3862 (1995). 
\item{[10]} Z. Liang and T. Meng,
             Z. Phys. {\bf A344}, 171 (1992),
             Phys. Rev. {\bf D49}, 3759 (1994);
 C. Boros, Z. Liang and T. Meng,
             Phys. Rev. Lett. {\bf 70}, 1751 (1993),
             Phys. Rev. {\bf D51}, 4867 (1995),
             Phys. Rev. {\bf D54}, 4680 (1996);
 C. Boros and Z. Liang, 
             Phys. Rev. {\bf D53, R}2279 (1996).
\end{description}

\end{document}